\documentclass[fp,twocolumn]{jpsj3}
\usepackage{txfonts}
\usepackage{graphicx}% Include figure files
\usepackage{dcolumn}% Align table columns on decimal point
\usepackage{bm}% bold math
\usepackage{color}
\usepackage{ulem}

\title{Nonequilibrium Pump-Probe Photoexcitation\\
as a Tool
for Analyzing Unoccupied Equilibrium States
of Correlated Electrons}

\author{
Youhei Yamaji$^1$\thanks{yamaji@ap.t.u-tokyo.ac.jp} and Masatoshi Imada$^2$
}
\inst{$^1$Quantum-Phase Electronics Center (QPEC), The University of Tokyo, Hongo, Bunkyo-ku, Tokyo, 113-8656, Japan \\
$^2$Department of Applied Physics, The University of Tokyo, Hongo, Bunkyo-ku, Tokyo, 113-8656, Japan
} %\\

\abst{
Relaxation of electrons in a Hubbard model coupled to
a dissipative bosonic bath is studied to simulate the
pump-probe photoemission measurement.
From this insight, we propose an experimental method of eliciting unoccupied part of the single-particle spectra at the equilibrium of doped-Mott insulators.  
We reveal first that effective temperatures of distribution functions
and electronic
spectra
are different during the relaxation, which makes the frequently employed thermalization picture inappropriate. 
Contrary to the conventional analysis, we show that the unoccupied spectra at equilibrium can be detected as the states
that relax faster.
}

\begin{document}
\newcommand{\bR}{\mbox{\boldmath $R$}}
\newcommand{\trr}[1]{\textcolor{black}{#1}}
\newcommand{\trs}[1]{\textcolor{red}{\sout{#1}}}
\newcommand{\tb}[1]{\textcolor{blue}{#1}}
\newcommand{\tbs}[1]{\textcolor{blue}{\sout{#1}}}
\newcommand{\mg}[1]{\textcolor{black}{#1}}
\newcommand{\tm}[1]{\textcolor{black}{#1}}
\newcommand{\tms}[1]{\textcolor{magenta}{\sout{#1}}}
\newcommand{\Ha}{\mathcal{H}}
\newcommand{\mh}{\mathsf{h}}
\newcommand{\mA}{\mathsf{A}}
\newcommand{\mB}{\mathsf{B}}
\newcommand{\mC}{\mathsf{C}}
\newcommand{\mS}{\mathsf{S}}
\newcommand{\mU}{\mathsf{U}}
\newcommand{\mX}{\mathsf{X}}
\newcommand{\sP}{\mathcal{P}}
\newcommand{\sL}{\mathcal{L}}
\newcommand{\sO}{\mathcal{O}}
\newcommand{\la}{\langle}
\newcommand{\ra}{\rangle}
\newcommand{\ga}{\alpha}
\newcommand{\gb}{\beta}
\newcommand{\gc}{\gamma}
\newcommand{\gs}{\sigma}
\newcommand{\vk}{{\bm{k}}}
\newcommand{\vq}{{\bm{q}}}
\newcommand{\vR}{{\bm{R}}}
\newcommand{\vQ}{{\bm{Q}}}
\newcommand{\vga}{{\bm{\alpha}}}
\newcommand{\vgc}{{\bm{\gamma}}}
\newcommand{\Ns}{N_{\text{s}}}
\newcommand{\avrg}[1]{\left\langle #1 \right\rangle}
\newcommand{\eqsa}[1]{\begin{eqnarray} #1 \end{eqnarray}}
\newcommand{\eqwd}[1]{\begin{widetext}\begin{eqnarray} #1 \end{eqnarray}\end{widetext}}
\newcommand{\hatd}[2]{\hat{ #1 }^{\dagger}_{ #2 }}
\newcommand{\hatn}[2]{\hat{ #1 }^{\ }_{ #2 }}
\newcommand{\wdtd}[2]{\widetilde{ #1 }^{\dagger}_{ #2 }}
\newcommand{\wdtn}[2]{\widetilde{ #1 }^{\ }_{ #2 }}
\newcommand{\cond}[1]{\overline{ #1 }_{0}}
\newcommand{\conp}[2]{\overline{ #1 }_{0#2}}
\newcommand{\nn}{\nonumber\\}
\newcommand{\cdt}{$\cdot$}
\newcommand{\bra}[1]{\langle#1|}
\newcommand{\ket}[1]{|#1\rangle}
\newcommand{\braket}[2]{\langle #1 | #2 \rangle}
\newcommand{\bvec}[1]{\mbox{\boldmath$#1$}}
\newcommand{\blue}[1]{{#1}}
\newcommand{\bl}[1]{{#1}}
\newcommand{\bn}[1]{\textcolor{blue}{#1}}
\newcommand{\rr}[1]{{#1}}
\newcommand{\bu}[1]{\textcolor{black}{#1}}
\newcommand{\but}[1]{\textcolor{blue}{#1}}
\newcommand{\cyan}[1]{\textcolor{black}{#1}}
\newcommand{\red}[1]{{#1}}
\newcommand{\fj}[1]{{#1}}
\newcommand{\green}[1]{{#1}}
\newcommand{\gr}[1]{\textcolor{green}{#1}}
\newcommand{\tg}[1]{\textcolor{black}{#1}}
\definecolor{green}{rgb}{0,0.5,0.1}
\definecolor{blue}{rgb}{0,0,0.8}

\maketitle

\section{Introduction}
Time-resolved and out-of-equilibrium
dynamics of electrons in \bu{pico to femtosecond time scale}
\textcolor{black}{has been revealed by rapidly developing pump-probe photoemission spectroscopy}~\cite{Bovensiepen}.
\trr{A challenge is to
get insight into emerging properties of  correlated electrons 
from the relaxing electrons far from equilibrium.}

\trr{Instead of
equilibrium Green's functions used to analyze
the single-particle excitations from the equilibrium~\cite{Damascelli},
nonequilibrium Green's functions have been used with a finite resolution limited by
laser pulse width~\cite{Freericks}.  However, understanding the full relaxation process including long-time scale remains a challenge.}

\trr{Electron relaxations 
were often analysed phenomenologically as if they were in equilibrium at each time steps but at
time-dependent
temperatures different from the state before pumping\textcolor{black}{~\cite{Freericks,
Allen,Perfetti,NJP_CDW,Moritz,TDARPES_CDW}}
such as two-temperature model\textcolor{black}{~\cite{Allen}}. 
However, we will show later that these quasithermal treatments are insufficient.}

\trr{In addition, the quasithermal approaches
do not tell us microscopic information
such as so-called dark sides (unoccupied sides) 
of the
spectra above the Fermi level~\cite{Reinert,Shimoyamada,Sakai}.
In fact, clarifying the \textcolor{black}{momentum-resolved}
\textcolor{black}{dark-side} spectra is claimed to hold a key for understanding challenging issues such as the mechanism of the pseudogap and high-$T_{\rm c}$ superconductivity in the cuprates~\cite{Sakai}. So far, the dark side has few accessibility such as the inverse photoemission with severely limited energy resolution.}

In this paper,
by introducing a simulation scheme for
nonequilibrium dynamics of many-body electrons, we first show that effective temperatures determined for the electronic spectra,
and the distribution (occupation) of electrons
follow time evolution clearly distinct each other, which unjustifies the quasithermal model. 
In addition,
we in fact find that electron relaxations strongly depend on excitations beyond the quasithermal picture.
This inhomogeneous relaxation emerges in a way that simpler electron-hole excitations relax faster.
We propose an experimental method by which unoccupied (dark side) quasiparticle states in equilibrium are identified by faster relaxation after pumping.

%{\it Hamiltonian.--}
\section{Model and Method}
\subsection{Hamiltonian}
\label{sub_ham}
Here, we simulate relaxation processes of
photo-excited states
of doped Mott insulators~\cite{ImadaFujimoriTokura}
by \bu{including}
phase relaxation process
in addition to energy, and momentum relaxations~\cite{Imry}.
We introduce a minimal model that consists
of a finite Hubbard \bu{ring} described by $\hat{H}_{\rm H}$ coupled to bosons, and 2-level atoms to understand the generic feature
of strongly correlated electrons in nonequilibrium.

\cyan{The energy and momentum
relaxations
of
electrons
in
crystalline solids occur
through electron-phonon interactions.
Simulating phonons
\bu{carrying}
finite momenta
are
\bu{necessary}
for the
\bu{relaxation.}
The phase relaxation is more complicated\bu{:}
\bu{In} a
closed system,
the phase relaxation of the
\bu{system}
never occur\bu{s}.
To simulate the phase relaxation,
we \bu{introduce} additional
two level atoms as detailed below.
\bu{\trr{At short intervals, we 
couple them to and disentangle them from}
the system consisting of electrons and phonons.
\textcolor{black}{When we disentangle these atoms,} 
we introduce
random phases and mimic the phase relaxation.}}

\begin{figure}[th]
\begin{center}
\includegraphics[width=9cm]{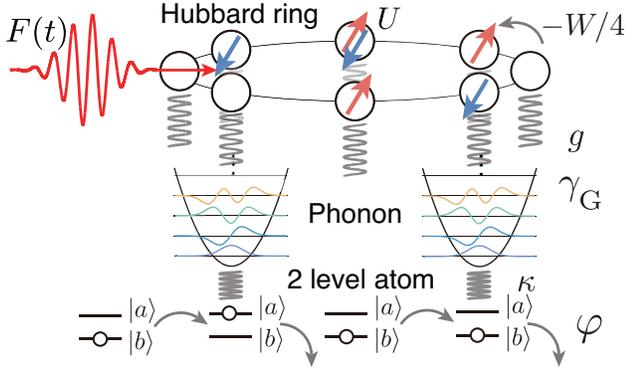}
\end{center}
\caption{(color online):
\cyan{Schematic description of our hamiltonian.
A Hubbard ring $\hat{H}_{\rm H} (t)$
coupled to
phonons $\hat{a}_{\pm q_0}$
is excited
by a time-dependent electric field $F(t)$.
These phonons are also coupled to two-level atoms.
We additionally introduce phonon dissipation with
a phenomenological decay constant $\gamma_{\rm G}$
%\sout{.}
\textcolor{black}{and with randomizing}
%\sout{By using a stochastic surrogate hamiltonian approach,
%we also randomize} 
the
%introduce phase relaxation due to random 
phases of wave function $\varphi$ (see Appendix \ref{subsub_swap} and \ref{subsub_Gisin} for details).
%\sout{The decay constant $\gamma_{\rm G}$ and the surrogate hamiltonian
%are detailed in Spplementary material.}
}
\label{fig_ham}} 
\end{figure}

Our hamiltonian
% \cyan{that minimally simulate energy, momentum, and phase relaxations}
reads
\eqsa{
  \hat{H}(t)=\hat{H}_{\rm H} (t) 
%+\hat{H}_{I} 
+ \hat{H}_{b} + \hat{H}_{\rm fb} + \hat{H}_{\rm JC},
}
\cyan{which is schematically illustrated in Fig. 1}.
%\textcolor{blue}{
A Hubbard \bu{ring under}
% under the influence of 
a time-dependent electric field $dF(t)/dt$
is defined as
%} 
\eqsa{
	\hat{H}_{\rm H} (t) = -\frac{W}{4}\sum_{i,\sigma}
						\left[
									e^{i F (t)} \hatd{c}{i+1\sigma}\hatn{c}{i\sigma}
									+ {\rm h.c.}
						\right]+U\sum_{i}\hatn{n}{i\uparrow}\hatn{n}{i\downarrow},
}
%and
%\eqsa{
%	\hat{H}_{I}
%						&=&
%						U\sum_{i}\hatn{n}{i\uparrow}\hatn{n}{i\downarrow},
%}
%\textcolor{blue}{
where, $W$ is the band width, $\hatd{c}{i\sigma}$ ($\hatn{c}{i\sigma}$) is
the operator creating (destroying) an electron with $\sigma$-spin at the $i$-th site,
and $U$ is the on-site Coulomb repulsion. Here the electron number operators are defined
as $\hatn{n}{i\sigma}=\hatd{c}{i\sigma}\hatn{c}{i\sigma}$.
%}

%\textcolor{blue}{
We introduce bosons that exchange energy and momentum 
to realize the energy, and momentum relaxations, 
which are subtantiated by (acoustic) phonons in real solids.  
%We study \tg{with} the finite Hubbard chain \textcolor{blue}{(described by $\hat{H}_0$ \tg{in} the following section)}.
%}
Two modes of bosons \mg{(denoted by $\nu$ and $\mu$)} are introduced by the Hamiltonian $\hat{H}_{\rm b}$
%as detailed in the following subsections\textcolor{blue}{ (
%\tg{The boson Hamiltonian $\hat{H}_{\rm b}$ is 
described by 
\eqsa{
  \hat{H}_{\rm b}&=&
\hbar\omega_{\rm b}\sum_{u=\nu,\mu}
  \left(\hatd{a}{u}\hatn{a}{u}+\frac{1}{2}\right),
}
where
$\hatd{a}{u}$ ($\hatn{a}{u}$) is the creation (annihilation) operator of a boson mode with the label $u=\nu,\mu$
with energy $\hbar\omega_{\rm b}$. Here $\nu=q_0$ and $\mu=-q_0$ specify momenta
%\sout{and momentum} 
$\pm q_{0}=\pm 2\pi /L$, where $L$ is the number of sites.
The coupling of the bosons to the electrons are described by $\hat{H}_{\rm fb}$:
\eqsa{
  \hat{H}_{\rm fb}&=&
	g\sum_{k=2\pi \ell /L}
	\left[
	\hatd{c}{k}\hatn{c}{k+q_{0}}
	\left(\hatn{a}{\mu}+\hatd{a}{\nu}\right)
	+
	%\hatd{c}{k+q_{0}}\hatn{c}{k}
	%\left(\hatn{a}{\nu}+\hatd{a}{\mu}\right)
	%\right]
  {\rm h.c.}
	\right],
}
%\textcolor{blue}{
where $g$ is the coupling constant.
% \tg{between} the electrons and bosons,
%\sout{\bu{and}
%$\hbar \omega_{\rm b}$ is the energy of the boson\bu{.}
%% modes.
%Here we define that
%$q_0 = 2\pi/L$,
%$\hatd{a}{\nu}=\hatd{a}{+q_{0}}$,
%and
%$\hatd{a}{\mu}=\hatd{a}{-q_{0}}$.}
These boson\bu{s}
% modes 
change the total momentum of the electron wave function
with minimal increment and decrement in momenta for the $L$-site Hubbard ring.
%}
We also introduce a 2-level atom \mg{coupled to} each mode of the boson.
These atoms absorb and supply the bosons. The coupling between the two 2-level atoms (labeled by $\alpha$ and $\beta$)
and bosons is described by  $\hat{H}_{\rm JC}$~\cite{Jaynes,Eberly} as 
\eqsa{
	\hat{H}_{\rm JC}
	=
 %\bu{2}
	\frac{\hbar
\omega_{\rm b}}{2}(\sigma_{z}^{(\alpha)}+\sigma_{z}^{(\beta)})
+\hbar \kappa \sum_{u=\nu,\mu}
 (\hat{a}^{\dagger}_{u}\sigma_{-}^{(\alpha)}+\hat{a}^{\ }_{u}\sigma_{+}^{(\alpha)}),
%	\nn
%	&+&\hbar \kappa (\hat{a}^{\dagger}_{\nu}\sigma_{-}^{(\alpha)}+\hat{a}^{\ }_{\nu}\sigma_{+}^{(\alpha)})
%	\nn
%	&+&\hbar \kappa (\hat{a}^{\dagger}_{\mu}\sigma_{-}^{(\beta)}+\hat{a}^{\ }_{\mu}\sigma_{+}^{(\beta)}),
}
%\textcolor{blue}{
where \bu{$\sigma^{(\alpha/\beta)}_z$ is a diagonal Pauli matrix and
$\sigma^{(\alpha/\beta)}_{\pm}$ are ladder operators interchanging
the eigenstates of $\sigma^{(\alpha/\beta)}_z$.} 
\if0
\eqsa{
\sigma^{(\alpha/\beta)}_z
&=&
\left[
\begin{array}{cc}
+1/2 & 0 \\
0 & -1/2 \\
\end{array}
\right],
\nonumber\\
\sigma^{(\alpha/\beta)}_{+}
&=&
\left[
\begin{array}{cc}
0 & 1 \\
0 & 0 \\
\end{array}
\right],
\nonumber\\
\sigma^{(\alpha/\beta)}_{-}
&=&
\left[
\begin{array}{cc}
0 & 0 \\
1 & 0 \\
\end{array}
\right].
\nonumber
}
\fi
$\hbar\kappa$ is the coupling constant between the 2-level atoms and bosons.
Here the energy difference
of the levels 
between the ground state and the excited state of the atom, denoted by
$\ket{b}$ and $\ket{a}$, respectively,
%of the atom 
is chosen as the energy of the boson mode $\hbar\omega_{\rm b}$.
%}

%Our time evolution of the finite Hubbard chain is governed by the time-dependent hamiltonian $\hat{H}(t)$  defined below.

%
By employing the time-dependent hamiltonian $\hat{H}(t)$,
the photo-excitation and relaxation processes of the correlated electron systems
will be studied in the following part of the present paper.
Before going into details, we remark the generality of the present one-dimensional model as a prototype of strongly correlated electron systems including those in higher dimensional systems.
%Here, a doubt about generality of the present hamiltonian may arise as a typical strongly correlated electron system:
One may suspect that the Tomonaga-Luttinger liquid behaviors may govern
dynamical processes of the one-dimensional Hubbard ring
which should be in a strict sense distinct from the universal Landau's Fermi liquid behaviors in higher dimensional correlated metals.
However, finite-size systems with finite time span and finite energy resolution in the present study do not distinguish 
the Tomonaga-Luttinger liquid behaviors from those of the Fermi liquid and do not hamper us from extracting
general tendency of the dynamical processes in the correlated electron systems. 

The relaxation of the dark-side spectra is
proposed to be dominated by the relaxation of simple particle-hole excitations.
%However, in the one-dimensional Hubbard ring employed in the present paper,
%the particle-hole excitations
% in the one-dimensional electron
%systems 
%As well-know facts, 
The particle-hole excitations in the one-dimensional electron systems
hardly remain as well-defined qausiparticles under the presence of the electron-electron interactions: 
The particle-hole excitations in one-dimensional interacting electron systems, for example,
described by the Hubbard ring are
reconstructed into collective excitations orthogonal to fermionic excitations in the thermodynamic limit,
which is characterized by the Tomonaga-Luttinger liquid behaviors.
However, the quasiparticle residue $Z$ remains finite even in the one-dimensional systems provided that the Hubbard ring with finite sites are examined:
The quasiparticle residue $Z$ in the one-dimensional interacting electron systems is scaled by the power of the system size
although $Z$ approaches zero as the system size increases\cite{Shashi}. 
Therefore, even though the one-dimensional Hubbard ring is employed,
the relaxation processes in the numerical simulation given in the following sections
are not specific to the one-dimensional systems except the relatively slow spin relaxations nearly decoupled from the charge excitations.

\subsection{Method for time evolution}
The wave function of the Hubbard ring coupled
to the bath can be expanded by using
the basis representing the $N$-electron eigenstates of the Hubbard ring $\ket{\Phi_m^{(N)}}$,
the Fock states of the phonons $\ket{n_{\nu}}$ and $\ket{n_{\mu}}$,
where $n_{\nu}$ and $n_{\mu}$ are numbers of phonons in the each modes,
and the eigenstates of the two-level atoms $\ket{\alpha}$ and $\ket{\beta}$,
as given as,
\eqsa{
 &&\ket{\Phi (t)}
 =
   \sum_{m=1}^{N_{\rm dim}}
   \sum_{n_{\nu}=0}^{L_{\rm B}-1}
   \sum_{n_{\mu}=0}^{L_{\rm B}-1}
   \sum_{\alpha=a,b}
   \sum_{\beta=a,b}
   C_{m;n_{\nu},n_{\mu};\alpha,\beta}(t)
   \nn
   &&\times
   \ket{\Phi_m^{(N)}}\otimes\ket{n_{\nu}}\otimes\ket{n_{\mu}}\otimes
   \ket{\alpha}\otimes\ket{\beta}.
}
As detailed below, the time-evolution based on the wave function
is twofold.
Together with the ordinary time-ordered hamiltonian dynamics
\eqsa{
 \ket{\Phi(t)}
 =
 T\exp
 \left[
 -i\int_{t_0}^{t}dt \hat{H}(t)
 \right]
 \ket{\Phi(t_{0})},
}
started with
%for 
the ground state wavefunction $\ket{\Phi(t_{0})}$,
%and the energy-momentum and phase
%relaxation introduced by periodic application of the swapping/partial swapping operators
%and the supplementary
%Gisin-type dissipation operator defined in the following subsection.
%\but{indispensable} 
energy, momentum and phase relaxations of the total system
are phenomenologically achieved through
the stochastic surrogate hamiltonian approach~\cite{Katz}
and
a damping constant $\gamma_{\rm G}$
of the boson numbers~\cite{Gisin}, as detailed in Appendix \ref{sub_relax} and \ref{sub_time}.
The former realizes
repeating and successive decoupling of the 2-level atoms from the bosons, and reset of the 2-level atoms to their ground states accompanied by randomization of \mg{their phases}, everytime
after a fixed interval of the time evolution by the Hamiltonan $\hat{H}$.
%\textcolor{red}{(see Appendix $\ast$ for details)}.
%\mg{This} is known as the stochastic surrogate hamiltonian approach\cite{Katz}. 
%\sout{We describe our surrogate Hamiltonain approach~\cite{SM}
% \but{\sout{and the \mg{time evolution} method}~\cite{SM}}
% \mgs{for the time evolution} 
%in \textcolor{black}{the Supplemental Material}}.
%If we just turn off the coupling between the atoms and the residual system \textcolor{blue}{(in other word, turn off $\hat{H}_{\rm JC}$)}, the entanglement between the atoms and the residual system remains.
%Therefore, when we add a set of new atoms and turn on the coupling, we need to expand the Hilbert space of the total wave function.
%To avoid the expansion and/or to keep the system size tractable, we employ the swapping operator of the stochastic surrogate hamiltonian approach\cite{Katz}. 

%\subsection{Hamiltonian}
%\gr{Many variables are introduced without definitions. Define $c, c^{\dagger} W,U,$ etc. etc.
%Also explain physical meaning of each term of the hamiltonian.}

We use a specific form of $F(t)$, 
\eqsa{
F(t)=
\int_{t_0}^{t} dt' A_{0} e^{-(t'-t_0 )^2 / t_{d}^{2}} \cos
\left[
\omega_{\rm pulse}(t'-t_{0})
\right],
}
to represent applied pulsed lasers 
%\textcolor{blue}{
with the amplitude $A_0$, the pulse center $t_0$, the pulse width $t_d$,
and the frequency $\omega_{\rm pulse}$.
%}.

\if0
%move the paragraph to Appendix
Then the wave function of the system is expanded by products of one electron's, two \tg{bosons', and two atoms'}
wave functions ($\ket{j}$, $\ket{n_{\nu}}$, $\ket{n_{\mu}}$, $\ket{\alpha}$, and $\ket{\beta}$, respectively) as,
\eqsa{
   \ket{\Phi (t)}
   &=&
   \sum_{j=1}^{N_{\rm dim}}
   \sum_{n_{\nu}=0}^{L_{\rm B}-1}
   \sum_{n_{\mu}=0}^{L_{\rm B}-1}
   \sum_{\alpha=a,b}
   \sum_{\beta=a,b}
   C_{j;n_{\nu},n_{\mu};\alpha,\beta}(t)
   \nn
   &\times&
   \ket{j}\otimes\ket{n_{\nu}}\otimes\ket{n_{\mu}}\otimes
   \ket{\alpha}\otimes\ket{\beta}.
}

The atoms have two levels as
\eqsa{
\ket{\alpha}, \ket{\beta}
=
\ket{a}, \ket{b},
}
where $\ket{a}$ is the excited state and $\ket{b}$ is the ground state of the atom.
\fi

\subsection{Spectral functions}
We calculate the momentum $k$ and frequency $\omega$ dependent spectral weight $A(k,\omega;t)$ \mg{and its occupied part $A_{\rm occ}(k,\omega; t)$} from
the time evolution,
\bu{by projecting the bosons and the two-level atoms out and} 
by taking
% \sout{the} 
\cyan{a} long-time approximation
% called
\bu{that we call}
%\sout{ the}
\cyan{stretched-time Lehmann-representation methods.}
The stretched-time Lehmann representation
is nothing but a Lehmann representation based on a {\it projected} electronic wave function $\ket{\Phi(t)}_{\rm proj}$
as detailed in Appendix \ref{sub_Ansatz}.
The projected wave function is prepared by operating a projection operator
\eqsa{
\hat{P}_0
&=&
(\sum_{m} \ket{\Phi_m^{(N)}}\bra{\Phi_m^{(N)}})
\nonumber\\
&\otimes&
\bra{n_{\nu}{\rm =0}}\otimes
\bra{n_{\mu}{\rm =0}}\otimes
\bra{{\rm \alpha = {\it b}}}\otimes
\bra{{\rm \beta  = {\it b}}},\label{prj_P0}
}
to the wave function $\ket{\Phi (t)}$
as
\eqsa{
\ket{\Phi(t)}_{\rm proj}
=
\hat{P}_0\ket{\Phi (t)}
=
\sum_{m=1}^{N_{\rm dim}}a_{m}(t)
\ket{\Phi_m^{(N)}},\label{prj_P1}
}
where $a_m (t)=C_{m;0,0;b,b}(t)$.

This method enables the frequency representation of transient phenomena
\bu{although it satisfies Freericks-Krishnamurthy-Pruschke's resolution
limit only approximately because of the finite probe-pulse width~\cite{Freericks}.
Typically $\mathcal{O}(10^2)$ fs pulses give us the spectrum resolution $\sim$ $\mathcal{O}(10^{-2})$ eV,
which justifies our results given below}.

%\noindent
%\subsection{Pumping}
%{\it Effective temperatures.--}
\subsection{Effective temperatures}
Here we show how to estimate the effective temperatures
of the physical quantities.
Our estimation is based on results of the angle-integrated spectra,
\eqsa{
A_{\rm occ}(\omega,t)=\sum_{k}A_{\rm occ}(k,\omega;t),
}
and
\eqsa{
A(\omega,t)=\sum_{k}A(k,\omega;t).
}
The distribution function is defined as
\eqsa{
n (\omega, t)=A_{\rm occ}(\omega,t)/A(\omega,t),
}
which corresponds to \bu{the Fermi-Dirac distribution if an effective temperature
% $T_{\rm eff}(t)$ 
becomes well-defined.}
% $f_{\rm th}(\omega, t)$,
%if the interpretation by a thermal equilibrium state with an effective temperature is justified.

When we fit the time-dependent
spectra by an
exact electronic spectra at a finite temperature \bu{$T_{\rm eff}$},
we can estimate effective temperatures
from two different functions,
namely,
the temperature of $n (\omega, t)$ and that of $A(\omega,t)$.
\bu{Here, we introduce the exact electronic (occupied) spectra at finite temperatures $A^{\rm (e)}(\omega,k_{\rm B}T_{\rm eff})$
and the distribution functions $n^{\rm (e)}(\omega,k_{\rm B}T_{\rm eff})$.}
\if0
\tg{If we project the wave function $|\Phi(t)\rangle$ onto the fermionic degrees of freedom as $|\Phi(t)\rangle_{\rm proj}$ and expand it by the basis of the energy eigenstate $|\Phi(t)_m^{(N)}\rangle $ with the energy $E_{m}^{(N)}$ as $|\Phi(t)\rangle_{\rm proj}=\sum_m a_m(t) |\Phi(t)_m^{(N)}\rangle $ (see Supplemental Material for details), the effective temperature $T_{\rm eff}(t)$ if it exists is related to $a_m(t)$ as} 
\eqsa{
a_{m}(t)
=
\exp
 \left[
 -
  %\frac{
  E_{m}^{(N)}/
  %}{
  2k_{\rm B}T_{\rm eff}(t)
  %}
 \right].
 \label{a(t)}
}
\tg{In fact, the exact finite temperature
spectra $A^{({\rm e})}(k,\omega;k_{\rm B}T_{\rm eff}(t))$
at equilibrium temperature $T=T_{\rm eff}(t)$ is obtained from Eq.(\ref{a(t)}).}
The finite temperature distribution
is given as
\eqsa{
n^{({\rm e})}(\omega,k_{\rm B}T_{\rm eff}(t))
=
\frac{
\sum_{k}A^{({\rm e})}_{\rm occ}(k,\omega;k_{\rm B}T_{\rm eff}(t))}
{
\sum_{k}A^{({\rm e})}(k,\omega;k_{\rm B}T_{\rm eff}(t))
}.
}
\fi
Then we can estimate the temperatures estimated from the
\textcolor{black}{following function}
by minimizing
\textcolor{black}{
\eqsa{
\chi^{2}[f^{\rm (e)},f](t;{T_{\rm eff}})
 =
 \int_{\Lambda_1}^{\Lambda_2}
 d\omega
 \left|
 f^{({\rm e})}(\omega,k_{\rm B}{T_{\rm eff}})
 -
 f (\omega,t)
 \right|^2.
 \label{chiff} 
\nonumber\\
}
}
\if0
\eqsa{
 \chi^{2}_{n}(t;T_{\rm eff})
 =
 \int_{\Lambda_1}^{\Lambda_2}
 d\omega
 \left|
 n^{({\rm e})}(\omega,k_{\rm B}T_{\rm eff})
 -
 n (\omega,t)
 \right|^2 
 \label{chinsq}
}
and
\eqsa{
 \chi^{2}_{A}(t;T_{\rm eff})
 =
 \int_{\Lambda_3}^{\Lambda_4}
 d\omega
 \left|
 A^{({\rm e})}(\omega,k_{\rm B}T_{\rm eff})
 -
 A (\omega,t)
 \right|^2 .
 \label{chiAsq}
}
\fi

The effective temperature $T_{\rm eff}$ that minimizes
\eqsa{
\chi^{2}_{n}(t;T_{\rm eff})\equiv \chi^{2}[n^{\rm (e)},n](t;T_{\rm eff}),
}
and
\eqsa{
\chi^{2}_{A}(t;T_{\rm eff})\equiv \chi^{2}[A^{\rm (e)},A](t;T_{\rm eff}),
}
are denoted by ${T_{\rm eff}}_{n}(t)$ and ${T_{\rm eff}}_{A}(t)$, respectively.
If the photo-excited states are really
described by the thermally excited ones,
these two temperatures should be the same,
namely, ${T_{\rm eff}}_{n}(t)={T_{\rm eff}}_{A}(t)$.

%\noindent
%\section
%{\it Results of simulations.--}
%\cyan{
\section{Results}
We simulate relaxation of photo-excited strongly correlated metals.
The spectra of the strongly correlated metals in equilibrium show
a typical structure regardless of dimensionality of the systems:
The spectra consist of the incoherent upper Hubbard band located around $\omega\sim U$,
the incoherent lower Hubbard band located for $\omega\lesssim 0$, and the low-energy coherent band
around the Fermi energy $\mu_{\rm F}$~\cite{Meinders,Senechal00,Kohno14}.
While the upper Hubbard continuum is well separated from the low-energy coherent band,
the latter and the lower Hubbard band touch each other or
are not separable, in general.
For the later discussion, we call the unoccupied part of the low-energy excitations including the quasiparticle band as mid-gap states.
We especially highlight
%show  
\tg{relaxations} of dark side mid-gap states
of doped Mott insulators after
photo excitation by a laser pulse.

Below, we set the energy unit \bu{$W/4=\hbar =1$}.
Parameters used for real-time evolutions are
$N_{\uparrow}=N_{\downarrow}=3$, $L=8$, $U=8$,
$\omega_{\rm b}=\sqrt{2}/4$, $g=0.05$, $\lambda=\tan^{-1}19\pi/40\sim 0.08$, and
\bu{$\gamma_{\rm G}\Delta t=0.05$ with a time interval $\Delta t=0.1$}.
\bu{We use pulses with frequencies $\omega_{\rm pulse}=\sqrt{2}/4$
much lower than direct gaps between occupied states and upper Hubbard bands.}
Typically, laser pulses used in this paper
give energy \cyan{$\sim 0.01$} mJ/cm$^2$ to electrons if we set $W/4=0.5$ eV
and the inter-site distance as \cyan{3} \AA.
%\sout{{\bf \tg{Realistic intersite distance should be, say, 3 \AA.}}}
The \tg{unit time of the present system corresponds to} $\sim$ \bu{1.32} fs.
%} 

%\cyan{
Here we use pulses with frequencies $\omega_{\rm pulse}$
much lower than direct gaps between occupied states and upper Hubbard bands.
Even though the photon energy is not enough for direct transitions to the upper Hubbard band,
%the non-equilibrium upper Hubbard bands appear.
both of the mid-gap low-energy states just above the Fermi level
and the upper Hubbard bands are excited \tg{due to the short width of pulses and the energy-time uncertainty relation.}
\if0
Even though direct transitions are prohibited,
pulsed laser excites electrons and induces a finite fraction of states at
nonequilibrium upper Hubbard bands, as detailed \tg{below}.
\fi
%}

%\noindent
%\subsection
\subsection{Relaxation of mid-gap states, upper Hubbard bands, and effective temperatures}
\begin{figure}[th]
\begin{center}
\includegraphics[width=8.5cm]{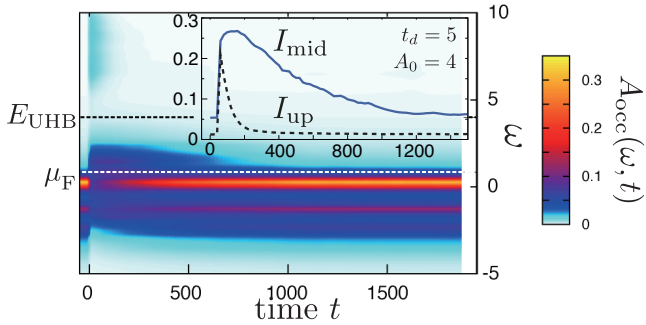}
\end{center}
\caption{(color online):
Relaxation of occupied density of states for $U/t=8$ with 6 electrons
($N_{\uparrow}=N_{\downarrow}$) on a 8-site ring.
A laser pulse with $t_d=5$, $A_0=4$, and $\omega_{\rm pulse}=\sqrt{2}/4$ is used.
The center of the pulse is located at $t=0$. 
\bu{Inset shows time-dependence of integrated spectral weights for mid-gap states and upper Hubbard bands,
$I_{\rm mid}$ and $I_{\rm up}$.
A finite line width $\delta=0.2$ is used for the spectra.}
\label{nequilibrium1}} 
\end{figure}

%\cyan{
First, we show relaxation of the density of states in Fig.\ref{nequilibrium1}.
A pulsed laser with duration $t_d=5$ and amplitude $A_0=4$ excites
the system and induces nonequilibrium mid-gap excited states
within an energy window, $\mu_{\rm F}\simeq 0.85<\omega\lesssim E_{\rm UHB}=4$,
%($\mu_{\rm F}\simeq 0.85<\omega\lesssim E_{\rm UHB}=4$),
%\but{[$\ast$There exists no clear meaning of $E=0$ for
%doped Mott insulators.]}, 
and upper Hubbard bands ($E_{\rm UHB}=4\lesssim \omega$),
where $\mu_{\rm F}$ is the Fermi energy.
Here, $E_{\rm UHB}$ is chosen as an upper bound
of the low-energy spectrum around the Fermi energy $\mu_{\rm F}$.

\tg{
We observe that the upper Hubbard band relaxes first.
The mid-gap states absorb the relaxed excitations from the upper Hubbard
band and it joins in the relaxation.
}   
%It is evident that the non-equilibrium upper Hubbard bands relax
%faster than the mid-gap states.
%}

%\cyan{
%As detailed below, the low energy excited states are not simply described by an effective electron temperatures. Spectral functions and distribution functions have different effective temperatures.
%}

%\cyan{
%The moderate relaxation time $\tr{T_{\rm eff}}_{\rm relax} \sim 3 \times 10^2$ enables us to detect the dark sides of the spectrum, which is detailed in the following subsections.
%}

\begin{figure}[h]
\begin{center}
\includegraphics[width=8.5cm]{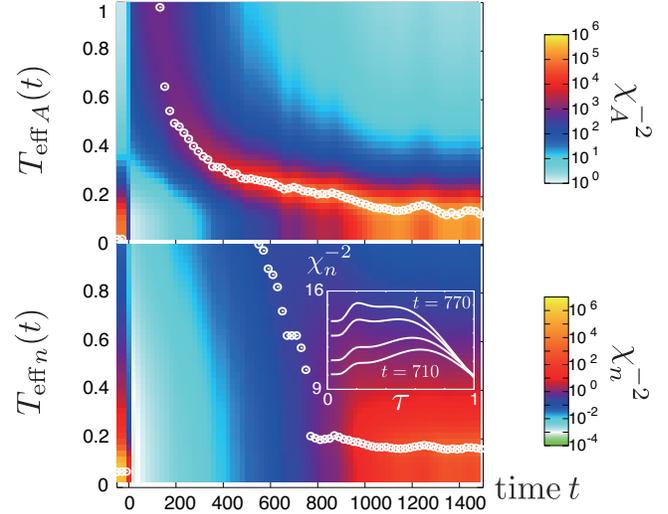}
\end{center}
\caption{(color online):
%[\textcolor{red}{$\ast$Temporal figure.
%$\ast\ast\ast$\tg{{\bf  Magnetic temperature is not defined yet.}\cyan{Magnetic
%correlation may be affected by one dimensionality.
%It is an artifact of the Hubbard ring.}}}]
\cyan{
Time-dependent effective temperatures of distribution functions ${T_{\rm eff}}_{n}(t)$ and
spectral functions ${T_{\rm eff}}_{A}(t)$ are shown by white circles with $\chi^{-2}_{n}$ and $\chi^{-2}_{A}$ in color plots.
\bu{Inset shows double peak structures of $\chi^{-2}_{n}(t;T_{\rm eff})$ around $t=740$, which induce
abrupt changes in ${T_{\rm eff}}_{n}(t)$.}
}
\label{nequilibrium_tmp}} 
\end{figure}
%With and without swapping.
%\cyan{
%To extract the non-equilibrium mid-gap states and upper Hubbard bands,

\textcolor{black}{By using the following integrated spectral weights,
\eqsa{
I_{E_1;E_2}(t)=\int_{E_1}^{E_2} d\omega
A_{\rm occ}(\omega,t)
/
\int_{-\infty}^{+\infty} d\omega
A_{\rm occ}(\omega,t),
}
we here define the integrated spectral weights for the mid-gap states and
upper Hubbard bands as $I_{\rm mid}(t)=I_{\mu_{\rm F};E_{\rm UHB}}(t)$ and
$I_{\rm up}(t)=I_{E_{\rm UHB};+\infty}(t)$, respectively.}
\bu{The} integrated spectral weights $I_{\rm mid}$ and $I_{\rm up}$
are summarized in the inset of Fig. \ref{nequilibrium1}.
\bu{The relaxation rate of $I_{\rm mid}$ up to $t\sim 10^3$
depends on the square of $g$
when other parameters are fixed, while the relaxation at longer time scales
is additionally affected by phase relaxation.
We employ $g$ to reproduce the experimental relaxation rate of doped Mott insulators~\cite{Perfetti,Nessler,Liu,Rameau} in the time range up to $t\sim 10^3$,
irrespective of ultrashort-time relaxation mechanisms
that may depend on dimensionality and details of the hamiltonian~\cite{Okamoto,Lenarcic,Aoki}.}

The numerical results of the effective temperatures ${T_{\rm eff}}_n (t)$ and ${T_{\rm eff}}_A (t)$
are shown in Fig.\ref{nequilibrium_tmp},
where the energy windows are chosen to extract
low-energy contribution as $\Lambda_1=-3$ and $\Lambda_2=+3$ ($\Lambda_1=-\infty$ and $\Lambda_2=E_{\rm UHB}$)
are set for $T_{{\rm eff}n}(t)$ ($T_{{\rm eff}A}(t)$) in Eq.(\ref{chiff}).
Around the occupied band bottom and the upper Hubbard band distant from the Fermi energy $\mu_{\rm F}$,
the density of states in non-equilibrium tends to deviate from the counterpart in equilibrium
more significantly than the density of states around the Fermi energy $\mu_{\rm F}$ (for example,
see Fig.1 of Ref.\citen{Ohnishi15}).

The effective temperatures ${T_{\rm eff}}_n (t)$ and ${T_{\rm eff}}_A (t)$
are significantly different as shown in Fig.\ref{nequilibrium_tmp}, which clearly demonstrates that
\cyan{the spectral weights relaxes faster than the distribution functions.
% during the exponential decay of the mid-gap states.
The deformation of the spectral weights is nothing but a manifestation of the electron correlations,
which invalidates
% the standard 
\bu{simple} effective electron temperature approaches.
The deformation of the spectral weights also means that the excited carriers above the Fermi level
are not injected to the dark-side of the equilibrium spectra, \mg{in contrast to}
% the effective temperature approaches
%and 
a naive expectation for visualizing the dark-side with hot carriers.}
%from time-dependent pump-probe measurements, as detailed in the following subsections.

In contrast to the relaxation of the spectral function,
the relaxation of the distribution function is not smooth as shown in the time dependence of ${T_{\rm eff}}_n (t)$
(see Fig.\ref{nequilibrium_tmp}).
The abrupt change in the time dependence of ${T_{\rm eff}}_n (t)$ is
a manifestation of a strong deviation from the equilibrium high-temperature distribution for a short time after the irradiation, $t\lesssim 800$. 
Similar strong deviations from the equilibrium distribution in initial stages of hot carrier relaxation are reported for
phonon-assisted relaxation of hot carriers in free electron systems~\cite{Ohnishi15}.

%\noindent
%\subsection
\subsection{Angle-resolved and unoccupied spectra through logarithmic differences in relaxation processes}
\label{AR_Unocc_LogDiff}
\begin{figure}[hbt]
\begin{center}
\includegraphics[width=8.0cm]{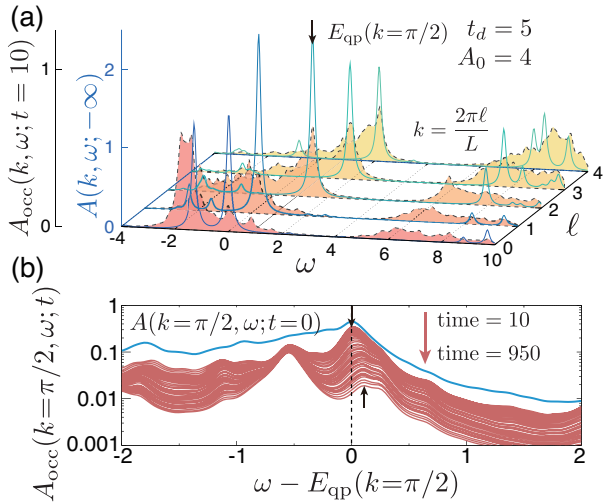}
\end{center}
\caption{(color online):
(a) Spectrum of a doped Mott insulator \mg{out of }equilibrium\cyan{, $A_{\rm occ}(k,\omega;t)$}
\cyan{in comparison with total spectrum in equilibrium $A(k,\omega;-\infty)$}.
The shaded area shows occupied spectrum $A_{\rm occ}(k,\omega;t)$ and
the solid curves show total spectrum $A(k,\omega;-\infty)$ \cyan{in equilibrium}.
%(b) Non-equilibrium spectrum just after the photo-excitation \tg{shown as shaded area}.
%The shaded area $A_{\rm occ}(k,w;t)$ appears different from the
%occupied states in equilibrium \tg{shown as solid curves}.
%\sout{Furthermore,} 
\cyan{The shaded area appears at energy ranges
where no equilibrium unoccupied states exist.} 
%{\bf \tg{Tiny weight should be more visible by improving the presentation.}}
(b) Time-dependent spectra at smallest unoccupied momentum $k=\pi/2$
from $t=10$ to $t=950$.
The nonequilibrium spectra have \tg{a prominent peak at the equilibrium quasiparticle peak $\omega=E_{\rm qp}
(k=\pi/2)$ (downward arrow) , while they have peaks
at $\omega$ different from the energy of the equilibrium peak as well}, as indicated by the upward vertical arrow.
\textcolor{black}{A finite line width $\delta=0.2$ is used as in Fig.\ref{nequilibrium1}.} 
\label{nequilibrium2}} 
\end{figure}

\tg{
Next, we examine momentum-resolved nonequilibrium spectra. 
%A naive expectation tells us that photo-excited electrons may occupy the unoccupied dark states similar to those in equilibrium.
%}
%\cyan{
In Fig. \ref{nequilibrium2}(a), we show the equilibrium momentum-resolved spectrum
with the same parameter set used in Fig. 2 and \ref{nequilibrium_tmp}.
Just after the photo-excitation, the occupied spectrum $A_{\rm occ}(k,w;t)$
becomes completely different from the spectrum in equilibrium
shown in Fig. \ref{nequilibrium2}(a).
As evident in Fig. \ref{nequilibrium2}(b), the peaks in nonequilibrium spectra
are not coincident with the peaks in equilibrium unoccupied states.
}

%\cyan{
Therefore, to observe or estimate the dark-side equilibrium spectra
of doped Mott insulators, photo-excited electrons are seemingly useless.
However, \tg{after closer inspection, we realize that,} from the photo-excited spectra, the dark-side equilibrium spectra
can be estimated as proposed below.  
%}

\begin{figure}[bh]
\begin{center}
\includegraphics[width=8cm]{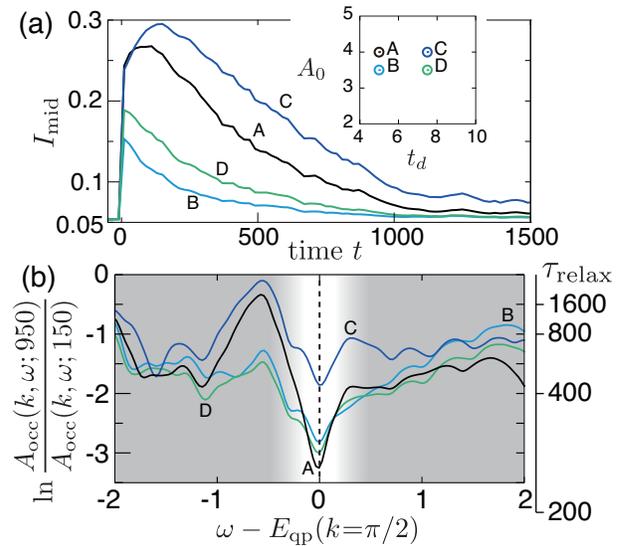}
\end{center}
\caption{(color online):
\cyan{(a) Time-dependences of
% integrated 
mid-gap
% spectral 
weights $I_{\rm mid}$.
The partial swap introduces noise in $I_{\rm mid}$.
Inset shows parameter sets of $t_d$ and $A_0$.
The
% black 
circle \bu{A} corresponds to the parameter set $t_d=5$ and $A_0=4$
%,
%which is i
used in Fig. 2 and 3.
(b) Logarithmic differences \bu{and $\tau_{\rm relax}$} in nonequilibrium occupied spectra at
the smallest unoccupied momentum $k=\pi/2$.   
%In (a) and (b), the same color data correspond to the same parameter set,
%shown in the inset of (a) with the symbol of the same color.
When laser pulses with $A_0 > 3$ are used,
the logarithmic differences show significant minima.
}
\label{nequilibrium3}} 
\end{figure}

%\noindent
%\subsection
%{\it Logarithmic differences.--}
%\cyan{
\cyan{Here, we propose an
% \mgs{alternative} 
approach \mg{to extract}
%\mgs{for visulalization of} 
the dark-side spectra
in equilibrium, which may replace the standard effective temperature approach. 
We} focus on energy- and momentum-resolved relaxation rates
of $A_{occ}(k,\omega;t)$ and frame a hypothesis:
%{\it 
%\sout{Excited nonequilibrium states have different relaxation rates
%and these relaxation rates}
{\it\textcolor{black}{The relaxation rates of excited nonequilibrium states} have local maxima when
the states \tg{are located} at the unoccupied quasiparticle energy in equilibrium.}
\cyan{The hypothesis seemingly contradicts Landau's Fermi liquid theory~\cite{AGD}, in which
the quasiparticles\bu{'
% are excitations with 
life times are} longer than
% other 
incoherent states in equilibrium.}
\cyan{However, outside the equilibrium, the quasiparticle excitations have efficient relaxation pathways as detailed below.}

\if0
\gr{This paragraph overlaps the description in Introduction}
\tbs{Before explaining our proposal based on the above hypothesis,
here, we emphasize potential generality of the present numerical results based on the Hubbard ring.
As it will be clear, the essential part of the hypothesis is summarized as follows:
If the nonequilibrium photo-excited states in the dark side consist of
single elementary excitations in equilibrium such as quasiparticle excitations in the Landau's Fermi liquids.
Even in the one-dimensional Hubbard ring, there are distinct elementary excitations as shown in Fig.\ref{nequilibrium2}.
In addition, the elementary excitations in the finite-size Hubbard ring have finite overlaps
with the quasiparticles in the Landau's Fermi liquid description as evident in finite quasiparticle residue $Z$
in the finite-size one-dimensional interacting electron systems.
Therefore, the elementary excitations in the present system are simply called as quasiparicles below. 
}
\fi

\cyan{Then, we explain how the photo-excited states in the dark side consisting
the quasiparticles relax faster than the incoherent excitations.
First of all,
we assume
% \sout{that}
\bu{correspondence of
a many-body excited state with small total momentum $|q_0|$
to
unoccupied quasiparticle peak due to added electrons in \textcolor{black}{equilibrium,} 
%\sout{the Lehmann representation
%of the ground state}, 
which is a central notion of the Fermi liquid theory.}
%single particle-hole excitations appear in the Lehmann representation as corresponding
%quasiparticle peaks, as in Landau's Fermi liquid theory, even in finite size systems.
In equilibrium, \mg{the} detailed balance \bu{between emission and absorption of the phonons
and the Pauli blocking}
% \mgs{, which} 
\bu{guarantee}
% same probabilities for gaining and losing momentum $q_0$ and
long life times of quasiparticles even in the presence of the electron-phonon couplings~\cite{Schrieffer}.
However, out of equilibrium, where \bu{the detailed balance and the Pauli blocking become inefficient},
single particle-hole excitation easily decays by emitting energy and momentum through the electron-phonon couplings.
%By taking into account 
\bu{Due to} the short-time violation of the energy conservation,
the particle-hole excitations decay through first order perturbation \bu{process of the}
%with respect to the
electron-phonon coupling $g$.
%\sout{For multicycle laser pulse, many-body excited states tend to have zero total momentum, and, thus,
%the excited states consist of a pair of particle-hole excitations with opposite momenta, which actually
%decay through the second order perturbation of $g$~\cite{SM}.} 
Therefore, by recalling the first assumption, out-of-equilibrium states injected at the quasiparticle peaks
decay faster than more complicated excited states such as multiple particle-hole excitations.}
%\textcolor{red}{[$\ast$ We may rationalize this hypothesis
%based on quasiparticles.]}
%}

%\cyan{
%Even when the hypothesis on the relaxation rates works, non-equilibrium excited states severely depends on the conditions of laser pulses, in general. Especially, energy distribution of excited states may strongly depend.
%\sout{To demonstrate the general validity of the hypothesis,}
%\sout{we choose several sets of parameters $t_d$ and $A_0$,} \sout{and
%show that the hypothesis works if the amplitude of the laser pulse
%is large enough \cyan{as illustrated in Fig.\ref{nequilibrium3}}.
%}
%\cyan{
To extract spectral components with large relaxation rates,
we propose and use logarithmic differences of nonequilibrium
occupied spectra $A_{\rm occ}(k,\omega;t)$ for some time interval from $t_1$ until $t_2$,
\bu{which gives an averaged relaxation time \textcolor{black}{$\tau_{\rm relax}(k,\omega)$} as}
%If the relaxation shows exponential decay with a relaxation time $\tau_{\rm relax}$,
%logarithmic differential gives us the inverse of \bu{$-\tau_{\rm relax}^{-1}
%\sim \frac{d}{dt}\ln A_{\rm occ} (k,\omega;t)$}.
%}
%\eqsa{
% -1/\tau_{\rm relax} \sim \frac{d}{dt}\ln A_{\rm occ} (k,\omega;t).
%}
%\cyan{
%When we can \tg{measure} spectral functions without any noise, the local minima of the logarithmic
%differential of $A_{\rm occ}(k,\omega;t)$ as functions of $\omega$ are nothing but the energy of the non-equilibrium states with large relaxation rates.
%}

%\cyan{
%However, in real pump-probe experiments, there is considerable noise. To get rid of the noise, we propose to integrate the logarithmic differential for some time interval \tg{in the region} where the relaxation becomes exponential, $t_1 < t < t_2$.
%Then, even in the presence of noise in spectra, the maxima of the relaxation rates of $A_{\rm occ}(k,\omega;t)$ as functions of $\omega$ are detectable by using the {\it logarithmic differences}:
%}
%\tg{Instead, to suppress the damage by unavoidable noise in experiments, we propose an average over some time interval as
\eqsa{
 \frac{-1}{\textcolor{black}{\tau_{\rm relax}(k,\omega)}}
\sim
\textcolor{black}{\frac{1}{t_2-t_1}}
%\int_{t_1}^{t_2}dt \frac{d}{dt}\ln A_{\rm occ}(k,\omega;t)=
\ln \frac{A_{\rm occ}(k,\omega;t_2)}{A_{\rm occ}(k,\omega;t_1)}.
}
\bu{By this average, the effects of unavoidable noise in experiments are suppressed.}
\bu{The slow relaxation of $I_{\rm mid}$ in comparison with the probe laser pulse width
($\sim$ 100 fs) validates our stretched-time Lehmann representation,
where \textcolor{black}{$t_2-t_1$ $(>0)$} longer than the relaxation time of $I_{\rm mid}$ is also better.}
%}

%\cyan{
In Fig.~\ref{nequilibrium3}, we show the \tg{usefulness} of the logarithmic differences of the
nonequilibrium unoccupied spectra.
For \tg{sufficiently} large amplitude of laser pulses, we obtain
significant minima of the logarithmic differences as
functions of $\omega$.
The minima agree with the quasiparticle peak\cyan{, as illustrated in Fig. \ref{nequilibrium3}(b)}. 
%atsmallest unoccupied momentum $E_{\rm qp}(k=\pi/2)$. 
In contrast, for smaller amplitude of laser pulses,
the minima become less significant and are not coincident with the quasiparticle peak
as shown in Fig.\ref{neuilibrium3} of Appendix \ref{sub_pump}.
%\sout{,
%as shown in Fig. \ref{nequilibrium3}(c).}

It is crucial to use the laser-pulse amplitude $A_0$
in an appropriate range.
% above a threshold.
% for efficiency of \bu{generating} single particle-hole excitations.
%To observe an unoccupied quasiparticle state with the momentum $q^*$, 
The laser pulse
must generate particle-hole excitations efficiently beyond the linear-response regime in a controlled way,
which requires $A_0$ to be comparable with that required for the Bloch
oscillation, $A_{\rm Bloch}$.
We propose to choose $A_0$ in an extended range roughly
given by $A_0 > \omega_{\rm pulsse}>1/t_d$ but not $A_0 \gg 1/t_d$.
See the detailed condition for the pump laser in the Appendix \ref{sub_pump}.

\section{Summary and Discussion}
In this paper, we have simulated
relaxation of photo-excited doped Mott insulators
by employing a Hubbard ring coupled with bosonic degrees of freedom.
We have shown the existence of two
distinct effective electronic temperatures that make the
standard effective-temperature approaches unjustified.
We instead propose a method to extract
the dark-side equilibrium spectra from nonequilibrium
pump-probe relaxation by using logarithmic difference of
the time-dependent spectra.
The logarithmic difference offers
a stable experimental tool that circumvents 
inevitable noises 
in the experiments.

%\textcolor{red}{
%To realize the present proposal based on the logarithmic difference of tdARPES in the time domain,
We have also explored sufficient conditions for detectable signals in experiments: Amplitude of the pumping laser is required to satisfy $A_0>\omega_{\rm pulse}> 1/t_{d}$.
%Pumping laser pulses with several sets of amplitude $A_0$ and duration $t_{d}$ have been numerically examined.
The laser pulses with a sufficiently large amplitude generates photo-excited states consisting of
combinations of elementary excitations and quasiparticle excitations. It 
enables us to extract the equilibrium dark-side spectra from the photo-excited nonequilibrium spectra.

%\textcolor{red}{
%Even though, due to the limitation of the present numerical method, 
%the hamiltonian studied in the present paper is a one-dimensional finite-size Hubbard ring coupled with bosonic baths,
The elementary excitations in equilibrium have relatively fast relaxation paths in 
the photo-excited nonequilibrium states.
The proposed method opens a novel way to access the momentum-resolved dark-side spectra of
the strongly correlated electron systems.
%In addition, the present proposal is based on a rather general observation
%that 
%Furthermore, as discussed in the section \ref{sub_ham} and \ref{AR_Unocc_LogDiff}, the one-dimensionality of the system
%may not hamper the generality of the present results for the finite-size system.
%Therefore, we believe that the essential physics discussed in the present paper
%is applicable to a wide range of strongly-correlated electron systems. 

The newly proposed scheme to access the dark-side spectra
will be helpful for the studies on unoccupied spectra of high-$T_{\rm c}$ superconducting cuprates~\cite{Sakai}.
It gives us a direct access to structure and formation of the so-called pseudogap in the cuprate superconductors,
which hold a key for understanding the mechanism of the high-temperature superconductivities in the cuprates.

\begin{acknowledgment}

%\acknowledgment
The authors thank Kyoko Ishizaka for fruitful discussion
about pump-probe measurements.
Y.Y. thank Takashi Oka and Yasuhiro Yamada for their comments on the present work.
We acknowledge the financial supports by a Grant-in-Aid
for Scientific Research (No. 22104010 and No. 22340090)
from MEXT, Japan, and a Grant-in-Aid for Scientific Research
on Innovative Areas `Materials Design through Computics:
Complex Correlation and Non-Equilibrium Dynamics.'
This work was also supported by MEXT HPCI Strategic
Programs for Innovative Research (SPIRE) (under the grant
number hp130007 and hp140215) and Computational Materials
Science Initiative (CMSI).

%For environments for acknowledgement(s) are available: \verb|acknowledgment|, \verb|acknowledgments|, \verb|acknowledgement|, and \verb|acknowledgements|.

\end{acknowledgment}

\appendix
In this Appendix, we detail the
simulation method for the energy, momentum and phase realized by the surrogated hamiltonian [17].
%\sout{ is described here}.
\textcolor{black}{The stretched-time Lehmann-representation methods are also introduced.}
\textcolor{black}{Constraint on pulse laser for efficient observation of quasiparticle excitations
is also detailed.}
\section{Energy, momentum, and phase relaxation}
\label{sub_relax}
\subsection{Swapping operator}
\label{subsub_swap}
As expalined \bu{in the main article},
we introduce an operator,
namely, swapping operator that stands for
the reset of the 2-level atoms to
its ground states $\ket{b}$ and
replacing the phases of the electronic and bosonic wave functions with random phases
\bu{with the periodic time interval $\Delta t$}.
Our swapping operator is defined as
\eqsa{
 &&\hat{O}_{\rm swap}\ket{\Phi (t)}
  =
   \sum_{m=1}^{N_{\rm dim}}
   \sum_{n_{\nu}=0}^{L_{\rm B}-1}
   \sum_{n_{\mu}=0}^{L_{\rm B}-1}
   \sqrt{
   \sum_{\alpha,\beta=a,b}
   \left|C_{m;n_{\nu},n_{\mu};\alpha,\beta}(t)\right|^{2}}
   \nn
   &&\times
   e^{i\varphi_{m;n_{\nu},n_{\mu}}}
   \ket{\Phi_m^{(N)}}\otimes\ket{n_{\nu}}\otimes\ket{n_{\mu}}\otimes
   \ket{b}\otimes\ket{b},
}
where $\varphi_{m;n_{\nu},n_{\mu}}$ is a random phase.

We also use a partial swap defined by the following partial swapping operator:
\eqsa{
 \hat{O}_{\rm pswap}=
 %\left[
 \hat{1}+
 \lambda\hat{O}_{\rm swap}
 %\right]
 ,
}
where $\hat{1}$ is an identity operator.
After applying the $\hat{O}_{\rm swap}$ and $\hat{O}_{\rm pswap}$,
we \mg{normalize} the wave function.

\subsection{Additional disspation mechanism}
\label{subsub_Gisin}
\textcolor{black}{To control the temperatures of the {\it lattice}, which is mimicked by
the two boson modes, in}
%In 
addition to the swapping operators,
we use a
% Gisin
Landau-Lifshitz and Gilbert-type dissipation operator to keep
the number of bosons small enough partially following Gisin's idea \textcolor{black}{(see Ref.\citen{Gisin})} 
\bu{for the time interval $\Delta t$}:
%\textcolor{red}{(we need the reference for the Gisin-type dissipation)},
\eqsa{
 &&\hat{O}_{\rm Gisin}\ket{\Phi (t)}
 =
   \sum_{m=1}^{N_{\rm dim}}
   \sum_{n_{\nu}=0}^{L_{\rm B}-1}
   \sum_{n_{\mu}=0}^{L_{\rm B}-1}
   \sum_{\alpha=a,b}
   \sum_{\beta=a,b}
   C_{m;n_{\nu},n_{\mu};\alpha,\beta}(t)
   \nn
   &&\times
   %\exp \left[
   e^{-\gamma_{\rm G}\Delta t (n_{\nu}+n_{\mu})}
   %\right]
   \ket{\Phi_m^{(N)}}\otimes\ket{n_{\nu}}\otimes\ket{n_{\mu}}\otimes
   \ket{\alpha}\otimes\ket{\beta},
}
where {$\gamma_{\rm G}$ is a phenomenological decay constant of the number of the bosons,}
%\sout{$\overline{n}$ is treated as a parameter \textcolor{blue}{ that controls the effective
%temperature of the bosons,} 
%which is self-consistently determined as the expectation value of the boson number
%in the original idea\cite{Gisin}}.
Here, we impose the condition that the equilibrium temperature 
to be zero, because experimental temperature is normally negligible in terms of the excited energy of electrons. 

\textcolor{black}{For practical numerical treatments, the additional dissipation mechanism
{with the decay constant $\gamma_{\rm G}$}
is helpful since it keeps the averaged bosonic number small and, therefore, reduces
the truncation errors for the bosonic Hilbert space.}
{In a real crystalline solids, the number of the phonon modes is
large enough to keep the density of the bosons.
However, due to the limit of computational resources,
we employ a minimal phonon subsystem.
Therefore, we need to get rid of {\it excess} accumulation of the bosons due to the limited number of the phonon modes.
The decay constant $\gamma_{\rm G}$ is again useful to avoid the accumulation of the bosons.}
%{\bf \gr{Explain in more detail why this additional dissipation is necessary.
%In particular explain why it can keep boson number small.
%Explanation on the relation to realistic electron systems is also useful. }}

\section{Time evolution}
\label{sub_time}
Our time evolution 
consists of 2 steps as detailed below:
First, we calculate the hamiltonian time-evolution
as
\eqsa{
 \ket{\widetilde{\Phi}(t_{i+1})}
 =
 T
 \exp
 \left[
 -i\int_{t_{i}}^{t_{i+1}}dt \hat{H}(t)
 \right]
 \ket{\Phi(t_{i})},
}
where $T
e^{
 -i\int_{t_{i}}^{t_{i+1}}\hat{H}(t)
}$ is a time-ordered operator.
Next, we apply the swapping operator,
\eqsa{
 \ket{\Phi(t_{i+1})}
 =
 \frac{
 \hat{O}_{\rm pswap}
 \hat{O}_{\rm Gisin}^{\textcolor{black}{M}}
 \ket{\widetilde{\Phi}(t_{i+1})}
 }
 {
 \sqrt{
 \left|
 \hat{O}_{\rm pswap}
 \hat{O}_{\rm Gisin}^{\textcolor{black}{M}}
 \ket{\widetilde{\Phi}(t_{i+1})}
 \right|^2
 }
 }.
}
We successively repeat the above 2 steps.
The swapping time defined by $t_{\rm swap}=t_{i+1}-t_{i}$ \textcolor{black}{$(=\textcolor{black}{M}\Delta t)$}
controls the phase relaxation, and an energy relaxation as well.
\textcolor{black}{In this study, we set $M=200$.}

\section{Ansatz for spectrum}
\label{sub_Ansatz}
%\textcolor{red}{We need to make a comment on the width of
%probing pulse.}
%
Here we introduce an approximate method to calculate
time-dependent spectral functions.  
The spectral functions revealed by time-dependent angle-resolved photoemission spectra provide us with 
physical pictures.
The present formalism corresponds to the limit of 
the long-time window of the laser pulse used as the probe \textcolor{black}{(see Ref.\citen{Freericks})}.
We call the following approximate method as \bu{{\it stretched-time Lehmann
spectral representation}.}

\subsection{Projection}
First, we need to eliminate the bosonic degrees of freedom and project
the entire wave function
\bu{onto} the $N$-electron partial Hilbert space \textcolor{black}{expanded by
$\{\ket{\Phi_{m}^{(N)}}\}$} as in Eq.(\ref{prj_P1}):
\eqsa{
% &&\ket{\Phi (t)}
% =
% \sum_{j,\alpha}\widetilde{\gamma}_{j\alpha}(t)\ket{\phi_{j}^{(N)}}\otimes\ket{\alpha}_{\rm env}
% \nn
% &&\rightarrow
% \nn
% &&
 \ket{\Phi (t)}_{\rm proj}=
\sum_{m=1}^{N_{\rm dim}}a_{m}(t)
\ket{\Phi_m^{(N)}}.
 %\ket{\alpha_0}
%_{\rm env}\bra{\alpha_0}
% \ket{\Phi (t)},
 %\sum_{i,\alpha}
 %\ket{\phi_{i}^{(N)}}\otimes\ket{\alpha}_{\rm env},
}
Here, we expand the projected wave function by the \bu{$N$-particle}
eigenstates of $\hat{H}_{\rm H}(t=0)$\textcolor{black}{, $\{\ket{\Phi_{m}^{(N)}}\}$}.
\if0
where 
$\ket{\alpha}_{\rm env}$ is the basis wave functions of the two bosons and the two two-level atoms, while the wave function $\ket{\alpha_{0}}_{\rm env}$ represents the ground states
of  bosons and atoms decoupled/isolated from the electrons.

After the projection, we expand the projected wave function by the \bu{$N$-particle}
eigen states of $\hat{H}_{\rm H}(t=0)$\textcolor{black}{, $\{\ket{\Phi_{m}^{(N)}}\}$} as
\eqsa{
 \ket{\Phi (t)}_{\rm proj}=\sum_{m} a_{m} (t) \ket{\Phi_{m}^{(N)}}.
}
\fi

\subsection{Ansatz}
\textcolor{black}{As an important observable in experiments, the time-dependent
spectral functions have often been discussed.
However, the energy-resolved spectral functions are definitely 
approximate ones whenever the observations are carried out in non-steady states \textcolor{black}{(see Ref.\citen{Freericks})}.
If the time-dependence of the system is slow enough in comparison with the time scale set by resolution limits,
the approximate time-dependence of the energy-resolved spectral functions becomes well-defined.
Here we apply the following ansatz justified in the steady-state limit:}
Starting from $\ket{\Phi (t)}_{\rm proj}$, we define an approximate
time-dependent Green's function for the occupied and unoccupied
states as, 
\eqsa{
  &&G_{\rm occ}(k,\omega; t)
  =
  \sum_{n,m}
  |a_m (t)|^2
  \nonumber\\
  &&\times
  \frac{
  \bra{\Phi^{(N)}_m}
  \hatd{c}{k\sigma}
  \ket{\Phi^{(N-1)}_n}
  \bra{\Phi^{(N-1)}_n}
  \hatn{c}{k\sigma}
  \ket{\Phi^{(N)}_m}
  }
  {
  \omega+i\delta+E_{n}^{(N-1)}-E_{m}^{(N)}
  },
%  \nn
%  &&=
%  \sum_{n, m}
%  \sum_{i_1 , i_2}
%  \sum_{j_1 , j_2}
%  |a_m (t)|^2
%  \frac{
%  \bra{\phi_{i_2}^{(N)}}
%  \hatd{c}{k\sigma}
%  \ket{\phi_{j_2}^{(N-1)}}
%  f_{i_2 m}^{\ast}
%  u_{j_2 n}
%  u_{j_1 n}^{\ast}
%  f_{i_1 m}
%  \bra{\phi_{j_1}^{(N-1)}}
%  \hatn{c}{k\sigma}\ket{\phi_{i_1}^{(N)}}
%  }
%  {
%  \omega+i\delta+E_{n}^{(N-1)}-E_{m}^{(N)}
%  }
%  \nn
%  &&=
%  \sum_{n,m}
%  |a_m (t)|^2
%  \frac{|g_{nm}^{(-)} (k\sigma)|^{2}}{
%  \omega+i\delta+E_{n}^{(N-1)}-E_{m}^{(N)}
%  }
 \label{Gocc}
}
and
\eqsa{
  &&G_{\rm uno}(k,\omega; t)=
  \sum_{n,m}
  |a_m (t)|^2
  \nonumber\\
  &&\times
  \frac{
  \bra{\Phi^{(N)}_m}
  \hatn{c}{k\sigma}
  \ket{\Phi^{(N+1)}_n}
  \bra{\Phi^{(N+1)}_n}
  \hatd{c}{k\sigma}
  \ket{\Phi^{(N)}_m}
  }
  {
  \omega+i\delta-E_{n}^{(N+1)}+E_{m}^{(N)}
  },
%  \nn
%  &=&
%  \sum_{n, m}
%  \sum_{i_1 , i_2}
%  \sum_{j_1 , j_2}
%  |a_m (t)|^2
%  \frac{
%  \bra{\phi_{i_2}^{(N)}}
%  \hatn{c}{k\sigma}
%  \ket{\phi_{j_2}^{(N+1)}}
%  f_{i_2 m}^{\ast}
%  v_{j_2 n}
%  v_{j_1 n}^{\ast}
%  f_{i_1 m}
%  \bra{\phi_{j_1}^{(N+1)}}
%  \hatd{c}{k\sigma}\ket{\phi_{i_1}^{(N)}}
%  }
%  {
%  \omega+i\delta-E_{n}^{(N+1)}+E_{m}^{(N)}
%  }
%  \nn
%  &&=
%  \sum_{n,m}
%  |a_m (t)|^2
%  \frac{|g_{nm}^{(+)} (k\sigma)|^{2}}{
%  \omega+i\delta-E_{n}^{(N+1)}+E_{m}^{(N)}
%  },
  \label{Guno}
}
where $E_m^{(N)}$ is the eigenvalue that corresponds to $\ket{\Phi_m^{(N)}}$.
\if0
where
matrix elements $g^{(\pm)}_{nm}(k\sigma)$ are
calculated as follows:
\eqsa{
 g_{nm}^{(-)} (k\sigma) = 
 \sum_{j,\ell}
 u_{\ell n}^{\ast}
 m_{\ell j}^{(-)}(k\sigma)
 f_{jm},
}
\eqsa{
 g_{nm}^{(+)} (k\sigma) = 
 \sum_{j,\ell}
 v_{\ell n}^{\ast}
 m_{\ell j}^{(+)}(k\sigma)
 f_{jm}.
}
%\gr{What is $f_{jm}$?}
Matrices $u$, $v$, $m^{(\pm)}$\textcolor{black}{, and $f_{jm}$} are determined
through the following expansions of the wave function
with $N$-particle Slater determinant with momentum
basis $\hat{c}_{k\sigma}$, $\phi_{i}^{(N)}$, as
\eqsa{
  \ket{\Phi^{(N)}_{m}}=\sum_{j}f_{jm} \ket{\phi_{j}^{(N)}},
}
\eqsa{
  \ket{\Phi^{(N-1)}_{n}}=\sum_{\ell}u_{\ell n} \ket{\phi_{\ell}^{(N-1)}},
}
\eqsa{
  \ket{\Phi^{(N+1)}_{n}}=\sum_{\ell}v_{\ell n} \ket{\phi_{\ell}^{(N+1)}},
}
and
\eqsa{
  m_{\ell j}^{(-)}(k\sigma)= 
  \bra{\phi_{\ell}^{(N-1)}}
  \hatn{c}{k\sigma}\ket{\phi_{j}^{(N)}},
}
\eqsa{
  m_{\ell j}^{(+)}(k\sigma)= 
  \bra{\phi_{\ell}^{(N+1)}}
  \hatd{c}{k\sigma}\ket{\phi_{j}^{(N)}}.
}

The projected wave function is also expanded by
the $N$-particle Slater determinants as
\eqsa{
 \ket{\Phi (t)}_{\rm proj}
 &=&
 \sum_{i}\gamma_{i}(t)\ket{\phi_{i}^{(N)}}
 \nn
 &=&
 \sum_{i}\gamma_{i}(t)\sum_{m}\ket{\Phi_{m}^{(N)}}\braket{\Phi_{m}^{(N)}}{\phi_{i}^{(N)}}
 \nn
 &=&
 \sum_{i,m}\gamma_{i}(t)f_{im}^{\ast} \ket{\Phi_{m}^{(N)}},
}
and the above expansion leads to
\eqsa{ 
 a_{m}(t)=\sum_{i}\gamma_{i}(t)f_{im}^{\ast}.
}
\fi

By using the above ansatz for the Green's functions, we can define
\bu{the
stretched-time Lehmann spectral representation}
% time-dependent spectral function 
as
\eqsa{
 A(k,\omega ; t)
 =-\frac{1}{\pi}{\rm Im}
 \left[
 G_{\rm occ}(k,\omega;t)
 +
 G_{\rm uno}(k,\omega;t)
 \right].
}
Here we note that the occupied spectrum is given by an occupied 
%\sout{spectral function} 
Green's function defined as
\eqsa{
 A_{\rm occ}(k,\omega ; t)
 =-\frac{1}{\pi}{\rm Im}
 \left[
 G_{\rm occ}(k,\omega;t)
 \right].
}

\section{Pumping laser dependence}
\label{sub_pump}
\begin{figure}[b]
\begin{center}
\includegraphics[width=7cm]{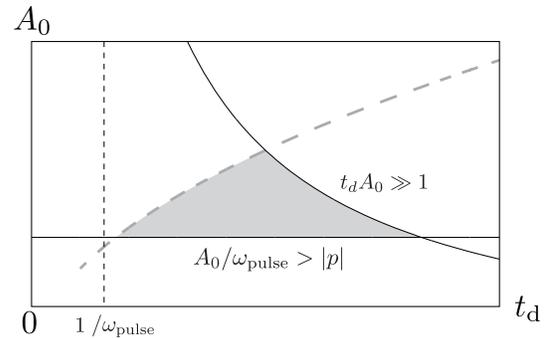}
\end{center}
\caption{
\textcolor{black}{Constraints on amplitude $A_0$ and duration $t_d$ of laser pulses
to observe quasiparticles with momentum amplitude $|p|$.
The curves and lines are determined by $A_0/\omega_{\rm pulse}=|p|$,
$t_d A_0 ={\rm const.}$,
and \textcolor{black}{$t_d=1/\omega_{\rm pulse}$}.
The dashed curve qualitatively illustrates the upper bound of $A_0$, \tm{below which the ``adiabatic transitions" accompanying the particle-hole excitations efficiently occur by suppressing the Landau-Zener tunneling.}
The shaded parameter region is suitable for generating a pair of
particle-hole excitations. }
\label{constraints}} 
\end{figure}

There is an optimal parameter range of amplitude and duration of laser pulses for
efficient observations of the dark-side spectra by utilizing
the logarithmic differences of the time-dependent angle-resolved photoemission intensities.
For the efficient observation, the laser pulses are required to efficiently
generate particle-hole excitations including the desired momentum $p$
{intended to measure the quasiparticle}.
{Below, we clarify what are the optimal amplitude and duration of the laser pulses.}

We start with two trivial constraints.
%There are two parameter regions where
{The efficient observations are hampered
in two regions.}
For \textcolor{black}{$\omega_{\rm pulse} t_{d} \ll 2\pi$}, the laser pulse contains
less than a single cycle
of electric-field oscillations, which is called a subcycle laser pulse.
Subcycle laser pulses are expected to generate net electric currents strongly depending
on their pulse shape \textcolor{black}{or phase}.
Thus, the subcycle pulses are not suitable for stable observations of the dark side.
The other trivial constraint comes from pump fluence.
For too large pump fluence or $t_d A_0 \gg 1$, the system gains too much energy.
Then, the relaxation processes involves multiple particle-hole excitations and
becomes complicated\textcolor{black}{, which may be observed as sudden increase in
relaxation time as laser fluence is increased}.

\begin{figure}[hb]
\begin{center}
\includegraphics[width=7.5cm]{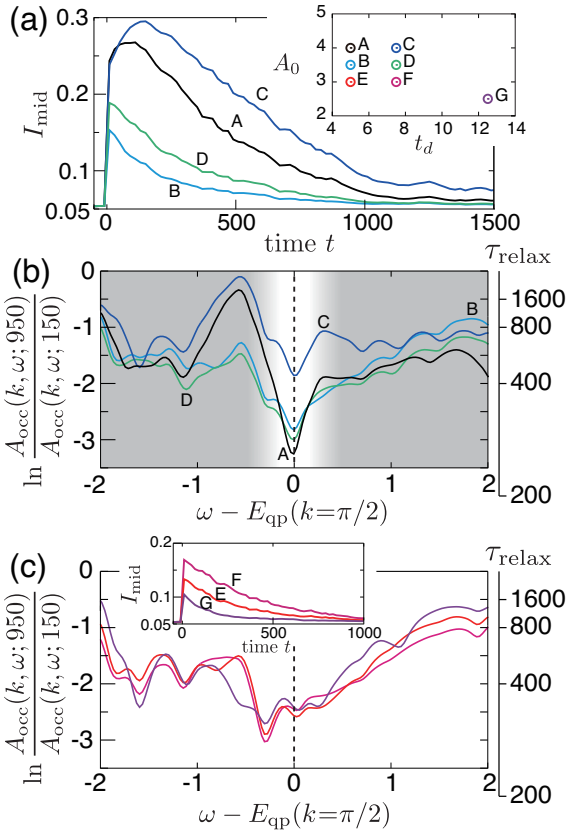}
\end{center}
\caption{
\textcolor{black}{(a) Time-dependences of
mid-gap
weights $I_{\rm mid}$.
The partial swap introduces noise in $I_{\rm mid}$.
Inset shows parameter sets of $t_d$ and $A_0$.
%The
%circle \bu{A} corresponds to the parameter set $t_d=5$ and $A_0=4$
%used in Fig. 2 and 3.
(b) Logarithmic differences \bu{and $\tau_{\rm relax}$} in nonequilibrium occupied spectra at
the smallest unoccupied momentum $k=\pi/2$.   
When laser pulses with $A_0 > 3$ are used,
the logarithmic differences show significant minima.
(c)
Logarithmic differences \bu{and $\tau_{\rm relax}$}
at
momentum $k=\pi/2$ for $A_0\lesssim 3$.
The minima are located at $\omega$ different from $E_{\rm qp}$.
\bu{Inset shows} $I_{\rm mid}$
for weakly excited cases with $A_0 \lesssim 3$.
}
\label{neuilibrium3}} 
\end{figure}
{To gain sufficient intensity of the emitted electrons by the probe laser, the pump light is required to generate sufficient amount of particle-hole excitations.
In the present formalism such excitations are generated by the acceleration of electrons by the time dependent vector potential
\textcolor{black}{$F(t)/ea$, where $e$ is the elementary charge and $a$ is the lattice constant.
The amplitude of the vector potential $F(t)/ea$ is controlled by a parameter $A_0$.}  
%probe Before going into details about other constraints,
%it is noteworthy that a single particle-hole excitation is hardly achieved in pumping processes.
%When multicycle pulse lasers with $\omega_{\rm pulse} t_d > 2\pi$ are used to pump correlated many-body electrons,
%such pulse lasers tend to generate many-body-excited states with zero total momenta.
%Therefore, the excited states consist of,
%at least, a pair of particle-hole excitations.

Let us clarify how \textcolor{black}{$F(t)\propto A_0$} generates the particle-hole excitations.
These particle-hole excited states orthogonal to the ground state
\textcolor{black}{in the electronic subsystem $\hat{H}_{\rm H}$}
are attained by many level crossings generated by the vector potential \textcolor{black}{$F(t) \propto A_0$} and additionally by the hybridization gap \tm{indirectly} induced by the coupling to bosons, which transforms the simple level crossing to the avoided crossing. 

\tm{When the system originally in the ground state is driven by adiabatically passing through such an avoided crossing, the state becomes containing particle-hole excitations originally orthogonal to the ground state in the absence of the light and bosons.} 
} 
Namely, the particle-hole excitations are generated by the ``adiabatic transitions" in the driven system.
\textcolor{black}{When the time dependent vector potential $F(t)\propto A_0$ is an oscillating
pulse with a frequency $\omega_{\rm pulse}$ and  a gaussian envelop $\exp \left[-(t-t_0)^2/t_d^2 \right]$,
$\omega_{\rm pulse}t_d$
determines the number of times that
% scales to how many times 
the system passes through the avoided crossing, where $t_0$ is the gaussian pulse center in the time domain and $t_d$ is the pulse width.}
\tm{Therefore, to increase the chance of such ``adiabatic transitions" to particle-hole excited states,
%it is better 
\textcolor{black}{we need}
to drive the system for sufficiently long time $t_d$
\textcolor{black}{at least to satisfy $\omega_{\rm pulse}t_d \gtrsim 1$} 
provided that the above condition $t_dA_0 \lesssim 1$ is satisfied.} 

\tm{\textcolor{black}{An additional requirement for the ``adiabatic transitions" to occur effectively exists:
%\sout{In addition,}Even if the condition $t_dA_0 \lesssim 1$ holds, a large $A_0$ leads to insufficient particle-hole
%generation, as follows: 
I}f $A_0$ becomes high, the probability of the non-adiabatic Landau-Zener tunneling at the avoided level crossings increases without following the adiabatic route. It prevents effective generation of particle-hole pairs.
Therefore, for higher $A_0$, $t_d$ is required to be longer to achieve the same density of particle-hole excitations. }

\tm{Equivalently, an appropriate density of particle-hole excitations are obtained in the region $(1-P_{\rm LZ}(A_0))\omega_{\rm pulse} \sim 1/t_d$, where the Landau-Zener tunneling probability $0<P_{\rm LZ}<1$ (the probability per each crossing through the avoided energy crossing point) depends on $A_0$ and largely increases with $A_0$. Thus the ``adiabatic transition" with the probability $1-P_{\rm LZ}(A_0)$ decreases with increasing $A_0$. This means that the upper limit of desired  $A_0$ increases with $t_d$.} 

\textcolor{black}{We also have a constraint that $A_0$ has to be larger than another threshold.
By noting that the time-dependent phase $F(t)$ causes lattice-momentum shifts,
for observation of quasiparticles with momentum amplitude $|p|$,
we need the maximum of the time-dependent phase $F(t)$ larger than $|p|$.
For a gaussian pulse, the maximum of the phase $F(t)$ is approximately given by $A_0/\omega_{\rm pulse}$.
Therefore, when we observe the quasiparticles with momentum amplitude $|p|$,
we need $A_0$ that satisfies $A_0/\omega_{\rm pulse} > |p|$.
In general, $A_0$ required by the above constraint is comparable with that
required for the Bloch oscillation, $A_{\rm Bloch}$.}
\if0
\tm{We also have a constraint that $A_0$ has to be larger than a\bn{nother} threshold.
Suppose $A_0$ is small. Then the adiabatic transition to excite the particle-hole excitation with the momentum $q$ may take place around the time $t$, \sout{where}\bn{when} $A_0t \sim  |q|$.
Therefore, the condition for the adiabatic transitions to occur is  $A_0/\omega_{\rm pulse} \bn{>}  |q|$ in one cycle $1/\omega_{\rm pulse}$.
Since the tunnelings occur in random directions, one needs $n=|p/q|^2$ cycles to excite the particle-hole excitation with momentum $p$. Then $A_0$ is required to satisfy
$A_0 \sqrt{t_d} > \sqrt{\omega_{\rm pulse}} |p|$, because $n\sim t_d\omega_{\rm pulse}$ holds. The tunneling probability depends on $g$ and also 
depends on $W/U$, which empirically reduces the threshold for $A_0$ slightly with increasing $W/U$.
In the finite-size calculation, the threshold is modified to
$A_0  > \sqrt{\omega_{\rm pulse}} |q_0|$, where $q_0$ is the minimum nonzero momentum, which appears in the present calculation but is a finite-size artifact.}
\fi
%where a hybridization gap is created by $g$. 
%There is a threshold for $A_0$ to approach an avoided level crossing where the Landau-Zener tunneling occurs.
%The threshold for $A_0$ is given by
%$f_{A}(W/U) A_0 /\omega_{\rm pulse}\sim \pi |q_0|/k_{\rm F}$, where $f_{A}$ is a monotonically-increasing dimensionless function
%with respect to $W/U$.
%The $W/U$ dependence of the factor $f_{A}$ is assumed based on
%the numerical simulations for larger $U/W$ than $U/(W/4)=8$.

By taking the above constraints into account, the suitable parameter region of
the pulse laser amplitude $A_0$ and duration $t_d$ is determined for the efficient
observation of the dark side.
The constraints and the obtained parameter region are summarized in Fig.\ref{constraints}.
\textcolor{black}{The appropriate region for the parameter values is roughly given by $A_0>\omega_{\rm pulse}
|p|\sim \omega_{\rm pulse} \gtrsim 1/t_d$.
The parameter values studied in the main text, namely, $A_0=4, t_d=5$, $\omega_{\rm pulse}=\sqrt{2}/4$,
and $|p|=2\pi/L\sim 0.8$
satisfies this criteria.} 

\textcolor{black}{To demonstrate the general validity of the hypothesis on the relaxation rates in the logarithmic differences,
we choose several sets of parameters $t_d$ and $A_0$, in addition to Fig.5 of the present paper, and
show that the hypothesis works if the amplitude of the laser pulse
is large enough, here $A_0 \gtrsim 3$, as illustrated in Fig.\ref{neuilibrium3}.}

\end{document}